\documentclass[10pt]{article}
\usepackage[cp1251]{inputenc}
\usepackage[english]{babel}

\title{{\bf \Large Non-Minimal Cosmological Model in Modified \\Yang--Mills Theory}\\
{\normalsize ~~{\bf V.\,K. Shchigolev}\thanks{E-mail:
vkshch@yahoo.com},~~{\bf G.\,N. Orekhova
\thanks{E-mail: gnorehova@mail.ru}}}\\
{\small {\it Ulyanovsk State University, 42 L. Tolstoy Str.,
Ulyanovsk 432000, Russia}}\\
\vspace{2mm}
\small \begin{quote}{\bf Abstract} --  In the present paper, we
consider a model of non-minimal modified Yang-Mills (Y-M) theory
in the Friedmann-Robertson-Walker (FRW) cosmology, in which the
Y-M field couples to the scalar curvature through a function of
its first invariant. We show that cosmic acceleration can be
realized due to non-minimal gravitational coupling of the modified
Y-M theory. Besides general study, we consider in detail the case
of power-law coupling function. We derive the basic equations for
the cosmic scale factor in our model, and provide several examples
of their solutions.\\
\vspace{2,5mm}
{\bf PACS numbers}: 98.80.-k, 98.80.Es, 04.30.-w, 04.62.+v\\
{\bf Key words}: Cosmological Model, Non-Minimal Coupling,
Yang-Mills Fields, Accelerated Expansion.\\
\end{quote}}
\date{}

\begin{document}

\maketitle \vspace{-2.5cm}
\section {Introduction}
\qquad Present accelerated expansion of the universe is well
proved in many papers \cite{C1}-\cite{C8}. In order to explain so
unexpected behavior of our universe, one can modify the
gravitational theory \cite{C9}-\cite{C14}, or construct various
field models of so-called dark energy which equation of state
satisfies $\gamma = p/\rho< -1/3$. The most studied models
consider a canonical scalar field (quintessence)
\cite{C15}-\cite{C17}, a phantom field, that is a scalar field
with a negative sign of the kinetic term \cite{C18}-\cite{C21}, or
the combination of quintessence and phantom in a unified model
named quintom \cite{C22}-\cite{C25}. In such field dark energy
scenarios, the potential choice plays a central role in the
determination of the cosmological evolution.

The alternative approach to the problem of accelerated expansion
is the consideration of various modifications  of the gravity
theory. Among such modifications, the theories with non-minimal
coupling of a field to gravity are especially attractive
\cite{C26}-\cite{C32}. For instance, scalar tensor theories are
generalization of the minimally coupled scalar field theories in a
sense that here the scalar field is non-minimally coupled with the
gravity sector of the action i.e with the Ricci scalar $R$. In
non-minimal theories, the matter field participates in the
gravitational interaction, unlike its counterpart in the minimally
coupled case where it behaves as a non gravitational source
\cite{C33}-\cite{C36}.

In particular, the discovery of the accelerated expansion of the
Universe encourages further development of the Y-M theory,
including its non-minimal coupling to gravity. Numerous attempts
have been made to consider modified  Y-M theories in cosmology as
alternatives to the dark energy (see, e.g., \cite{C37} and
references therein). One of the directions in such a
generalization of the Y-M theory is connected with a non-minimal
extension of the Y-M field theory. There are several motivations
to study non-minimal Y-M theory (see, e.g., \cite{C38,C39} and
references therein).

Basically, the present study is a sequel to the paper \cite{C40},
where the modified Y-M theory is investigated under the condition
of minimal coupling to gravity. Following the considerations
provided in \cite{C40}, we study a model of non-minimal modified
Y-M theory, in which the Y-M field couples to a function of the
scalar curvature. We show that the accelerated expansion can be
realized due to the non-minimal gravitational coupling of Y-M
field. Besides general study, we give some examples of exact
solution for the model under consideration, which surely can not
cover all possible applications of this research.

\section {\large The model equations}

\quad Let the action of our model be presented by a generalization
of the modified Y-M action \cite{C40} on the case of non-minimal
coupling:
\begin{equation}
\label {1} S = -\int d^4x
\sqrt{-g}\Bigl\{\frac{R}{2\kappa}\Psi(F^a_{ik}F^{aik})+\Phi(F^a_{ik}F^{aik})\Bigr\},
\end{equation}
where the Yang-Mills tensor is
$F_{ik}^{a}=\partial_{i}W_{k}^{a}-\partial
_{k}W_{i}^{a}+f^{abc}W_{i}^{b}W_{k}^{c}$, and $\Psi$ and $\Phi$
are the arbitrary differentiable functions. In the case of
$\Psi\equiv1$, $\Phi
=\displaystyle\frac{1}{16\pi}F_{ik}^{a}F^{aik}$ , this action
describes the Einstein-Yang-Mills theory. We assume that the
universe space-time is described by a Friedmann-Robertson-Walker
(FRW) geometry:
\begin{equation}
\label {2} d s^2 = N(t)^2 d t^2- a^2 (t)(d r^2+\xi^2 (r)d \Omega
^2),
\end{equation}
where $\xi(r)=\sin r,r,\sinh r$ \ in accordance with a sign of the
curvature $k=+1,0,-1$. To study the model based on action
(\ref{1}) in metrics (\ref{2}), we substitute this metrics into
action (\ref{1}) and take into account that
$$
R=-6\frac{a\ddot{a}N-a\dot{a}\dot{N}+\dot{a}^{2}N}{a^{2}N^{3}}
$$
for this metrics. As a result, we can obtain the following
effective Lagrangian per unit solid angle:
\begin{equation}
L_{eff}=\frac{3}{8\pi G}\Bigl(\frac{a^{2}\ddot{a}}{N}+\frac{a\dot{a}^{2}}%
{N}-\frac{a^{2}\dot{a}\dot{N}}{N^{2}}+kaN\Bigr)\Psi(I)\xi^{2}-\Phi(I)a^{3}%
N\xi^{2},\label{3}
\end{equation}
where $I=F_{ik}^{a}F^{aik}$. At the same time, the generalized
Wu-Yang ansatz for $SO_{3}$ Yang-Mills field can be written down
as  \cite{C41}:
$$
W_{0}^{a}=x^{a}\frac{W(r,t)}{er},W_{\mu}^{a}=\varepsilon_{\mu
ab}x^{b}
\frac{K(r,t)-1}{er^{2}}+\Bigl(\delta_{\mu}^{a}-\frac{x^{a}x_{\mu}}{r^{2}
}\Bigr)\frac{S(r,t)}{er}.
$$
Substituting
$$
K(r,t)=P(r)\cos\alpha(t),~S(r,t)=P(r)\sin\alpha(t),~W(r,t)=\dot{\alpha}(t)
$$
into this ansatz, we get the following components of Y-M tensor
\cite{C41}
\begin{eqnarray}
{\bf F}_{01}={\bf F}_{02}={\bf F}_{03}=0,~~~~~~ {\bf
F}_{12}=e^{-1} P'(r)\Bigl({\bf m }\,\cos\alpha + {\bf
l}\,\sin\alpha\Bigr),\nonumber\\
{\bf F}_{13}=e^{-1} P'(r)\sin\theta\Bigl({\bf m }\,\sin\alpha-{\bf
l}\,\cos\alpha\Bigr),~~~ {\bf
F}_{23}=e^{-1}\sin\theta\Bigl(P^2(r)-1\Bigr){\bf n}, \label {4}
\end{eqnarray}
presented in the orthonormalized isoframe \thinspace\thinspace$\mathbf{n}%
=(\sin\theta\cos\phi,\sin\theta\sin\phi,\cos\theta),~~\mathbf{l}=(\cos
\theta\cos\phi,\cos\theta\sin\phi,-\sin\theta)$ and
$\mathbf{m}=(-\sin \phi,\cos\phi,0)$. The prime in (\ref{4})
stands for the derivative with respect to $r$. As it has been
noted in \cite{C41}, Y-M field (\ref{4}) possesses only magnetic
components. From formulas (\ref{2}) and (\ref{4}), it is easy to
find that the Y-M invariant $I=F_{ik}^{a}F^{aik}$ has the
following expression:
\begin{equation}
I=\frac{2}{e^{2}a^{4}\xi^{2}}\Bigl[2P^{\prime}{}^{2}+\frac{(P^{-}1)^{2}}
{\xi^{2}}\Bigr].\label{5}
\end{equation}
Varying the Lagrangian density (\ref{3}) over $P(r)$ and taking
into account (\ref{5}), we obtain the following Euler--Lagrange
equation for the Y-M field:
\begin{equation}
\Bigl\{P^{\prime\prime}-\frac{(P^{2}-1)P}{\xi^{2}}\Bigr\}\Bigl[Q(t)\Psi^{\prime
}+a^{3}\Phi^{\prime}\Bigr]+P^{\prime}\frac{\partial I}{\partial
r}\Bigl[Q(t)\Psi
^{\prime\prime}+a^{3}\Phi^{\prime\prime}\Bigr]=0,\label{6}
\end{equation}
where $\Phi^{\prime}\equiv d\Phi(I)/dI,\,\Psi^{\prime}\equiv
d\Psi(I)/dI$ and the following notation is temporarily introduced:
\[
Q(t)=\frac{3}{8\pi G}\Bigl(\frac{a^{2}\ddot{a}}{N}+\frac{a\dot{a}^{2}}%
{N}-\frac{a^{2}\dot{a}\dot{N}}{N^{2}}+kaN\Bigr).
\]

The particular solution of the Y-M equation in the FRW metrics has
been obtained  in \cite{C41}. It has the form $P(r)=
\xi^{\prime}(r)=\cos r, \cosh r$ for $k=+1,-1$ consequently. It
satisfies equation (\ref{6}) as well. Indeed, it turns the
expression in braces  into zero and satisfies, in view of
(\ref{5}), the equality $\displaystyle\frac{\partial I}{\partial
r}=0$. It is easy to prove the latter, as for this solution the
Y-M invariant depends only on time:
\begin{equation}
\label{7}I = I(t) = \frac{6}{e^{2} a^{4}(t)}.
\end{equation}
Varying Lagrangian over $a(t)$ and $N(t)$ with the subsequent
choice of gauge $N=1$,  one can obtain the following equations for
our model:
\begin{eqnarray}
\label {8} \Bigl[2\frac{\ddot a}{a}+\Bigl( \frac{\dot
a}{a}\Bigr)^2 + \frac{k}{a^2}\Bigr]\Psi(I)-4\, I\,
\Psi'(I)\Bigl[2\frac{\ddot a}{a}-2 \Bigl( \frac{\dot a}{a}\Bigr)^2
+ \frac{k}{a^2}\Big]+ 16\Bigl( \frac{\dot
a}{a}\Bigr)^2 I^2 \Psi''(I)=\nonumber\\
= 8\pi G \Big[\Phi(I)-\frac{4}{3}\, I\,\Phi'(I)\Bigr],
\end{eqnarray}

\begin{equation}
\label{9}\Big[\Bigl(\frac{\dot a}{a}\Bigr)^{2} + \frac{k}{a^{2}}%
\Big] \Psi(I)-4\Bigl(\frac{\dot a}{a}\Bigr)^{2} I\,
\Psi^{\prime}(I)= \frac{8\pi G}{3}\Phi(I),
\end{equation}

\section{\large Several examples of exact solution}

\quad\textbf{(I)} The arbitrariness of differentiable functions
$\Psi(I) $ and $\Phi(I) $ essentially  complicates the general
analysis of the model equations (\ref{8}),(\ref{9}). Therefore, we
will consider some special cases. Let us begin with the power-law
dependencies:
\begin{equation}
\label{10}\Phi(I)=A\,I^{n},~~~\Psi(I)=B\,I^{m},
\end{equation}
where $A$ and $B$ are some dimensional constants, free parameters
of the model. The substitution of expressions (\ref{10}) into
equations (\ref{8}), (\ref{9}) leads to the following set of
equations:
\begin{equation}
\label{11}(1-4m)\Bigl[2\frac{\ddot a}{a}+\Bigl( \frac{\dot
a}{a}\Bigr)^{2} (1-4m) + \frac{k}{a^{2}}\Bigr] = \frac{8\pi
G}{3}\, \frac{A}{B}(3-4n)\,I^{n-m} ,
\end{equation}
\begin{equation}
\label{12}\Bigl(\frac{\dot a}{a}\Bigr)^{2} (1-4m) +
\frac{k}{a^{2}}= \frac{8\pi G}{3}\,\frac{A}{B}\,\,I^{n-m}.
\end{equation}
As one can see from these equations, the case $m=1/4$ is a
specific one. Indeed,  from equation (\ref{11}) with $m=1/4$ it
immediately follows that $n=3/4$, i.e. the power of $I$ in
equation (\ref{12}) is equal to $n-m = 1/2$ , and the first term
in the left-hand-side of this equation is equal to zero. Taking
into account (\ref{7}), we can  reduce equation (\ref{12}) to a
ratio between constants $A $ and $B $ of the form: $\displaystyle
k =\frac{8\sqrt{6} \pi G} {3e} \frac{A} {B} $. Thus, the scale
factor remains uncertain, or arbitrary. It is easy to prove that
the same unclearly interpreted result will turn out, if functions
(\ref{10}) are directly substituted into Lagrangian (\ref{3})
together with $n=3/4, \,m=1/4$ (or $\displaystyle
\Phi=A(6/e^{2})^{3/4}a^{-3}(t),\,\Psi=A(6/e^{2})^{1/4}a^{-1}(t)$).
Varying then it over $a(t)$ and $N(t)$, we again arrive at the
same relationship between $A$ and $B$. As for the equation for the
second derivative of the scale factor, it will be satisfied
identically for any $a(t)$. Putting this case aside, we consider
our model with $m\ne1/4$.

Combining (\ref{11}) and (\ref{12}), we have the following
equation:
\begin{equation}
\label{13}\frac{\ddot a}{a} = \frac{8\pi G}{3} \frac{A}{B}\, \frac
{[2(n-m)-1]}{(4m-1)}\,\,I^{n-m}
\end{equation}
for the second derivative of the scale factor.

It follows from (\ref{7}) that  $I> 0$. Therefore, the necessary
condition of accelerated expansion of the universe ($\ddot a> 0$),
following from equation (\ref{13}), is reduced to
$$
\frac{A}{B}\,\frac{[2(n-m)-1]}{(4m-1)}>0.
$$
The latter inequality must be solved for $m>1/4$, or  $m<1/4$. The
result of solving is presented  in Table 1. \vspace{3mm}
\begin{center}
\begin{tabular}
[c]{|l||l|l|l|}\hline
\vspace{-3mm} \, & \, & \, & \,\\
Ia & $A/B>0$ &$m > 1/4$ & $n-m > 1/2$\\\hline
\vspace{-3mm} \, & \, & \, & \,\\
Ib & $A/B>0$ & $m < 1/4$ & $n-m < 1/2$\\\hline
\vspace{-3mm} \, & \, & \, & \,\\
Ic & $A/B<0$ & $m > 1/4$ & $n-m < 1/2$\\\hline
\vspace{-3mm} \, & \, & \, & \,\\
Id & $A/B<0$ & $m < 1/4$ & $n-m > 1/2$\\\hline
\end{tabular}

\end{center}
\begin{quote}
\textbf{Table 1.} Conditions for the free parameters $A/B, \, m$
and $n-m $, corresponding to corresponding to accelerated
expansion according to equation (\ref{13}).
\end{quote}
\vspace{3mm}

It is necessary to emphasize  that equation (\ref{13}) is employed
for the analysis of the accelerated mode of evolution, but is a
differential consequence of equation (\ref{12}). Therefore, the
model dynamics is defined by the only independent equation
(\ref{12}). As one aims to solve this equation at the condition of
accelerated expansion, it is necessary to take into account Table
1.

From Table 1, it follows that in the case of standard Y-M
Lagrangian ($n=1$) the accelerated expansion is possible, if (Ia)
\, $A/B> 0$ and $m \in(1/4,1/2) $, or (Id) \, $A/B <0$ and $m
<1/4$ are realized. The assumption of minimal coupling (that is
$m=0$) corresponds to (Id). However, equation (\ref{12}) at $A/B
<0, m=0, n=1$ and ($k = + 1$) has no the real solution, and the
case (Ia) does not correspond to $m=0$. The latter simply means
the absence of the accelerated mode in Einstein-Yang-Mills minimal
theory \cite{C37}. The trivial solution with accelerated expansion
can be obtained in the case $A/B <0, m=0, n=1$, and for the
negative sign of curvature. Taking into account (\ref{7}), and
supposing $A=1/16\pi, \, B =-1$,  we can express equation
(\ref{12}) and its solution as follows:
\begin{equation}
\label{14}\dot a^{2}=1-\frac{G}{e^{2}}\,\frac{1}{a^{2}},~~~~ a(t)=
\sqrt {\frac{G}{e^{2}}+ t^{2}},
\end{equation}
where the constant of integration is equal to zero for the sake of
simplicity. For this solution, the acceleration equals $\ddot a =
(G/e^{2})/[(G/e^{2})+t^{2}]^{3/2}>0$, and it decreases to zero
with time. It is interesting that this model of non-massive Y-M
field with linear dependance on $I = F^{a} _{ik} F ^{aik} $ leads
to the accelerated expansion. However, due to equation (\ref{12}),
this model violates the weak energy condition. Therefore this
example of solution (\ref{14}) is exclusively illustrative.

Let us consider now two examples of exact solution for this model
which are not so trivial but rather simple.

{\bf a}) Let $n = 1, \, m=3/4$, that is we will consider the case
(Ic) from Table 1. Then equation (\ref{12}) can be written down as
follows:
\[
2\,\dot a^{2}-k= C\,a,~~\mbox{where}~~C=\frac{8\pi G}{3}\left( \frac{6}{e^{2}%
}\right) ^{1/4}\,\Bigl|\frac{A}{B}\Bigr|.
\]
The obvious solution for this equation is:
$$
a(t)=\frac{C}{8}(t-C_{0})^{2}-\frac{k}{C},
$$
where $C_{0}^{2} \geq 8k/C^{2}$.  This model experiences constant
acceleration: $\ddot a = C/4$, and the Hubble
parameter is equal $\displaystyle H = \frac{2(t-C_{0})}{(t-C_{0})^{2}%
-8k/C^{2}}$.

{\bf b}) We now consider an example of exact solution with $n = m$
that corresponds to (Ib) and (Ic) in Table 1. In these cases,
equation (\ref{12}) can be written down as follows:
\begin{equation}
\label{15}M\,\dot a^{2}-\delta\,k= C\,a^{2},
\end{equation}
where
$$
C=\frac{8\pi G}{3}\,\Bigl|\frac{A}{B}\Bigr|,~~~M = | 4m-1 |,~~~
\delta= \left\{
\begin{array}
[c]{rcl}
+1 ~~~\mbox{for}~~m >1/4\, , &  & \\
-1~~~\mbox{for}~~m <1/4\, . &  &
\end{array}
\right.
$$

For the positive sing of curvature ($k = + 1$) and $m> 1/4$, as
well as for the negative sign of curvature ($k =-1$) and $m <1/4$,
the exact solution for equation (\ref{15}) is equal
$$
a(t)=
\frac{1}{\sqrt{C}}\sinh\Bigl(\sqrt{\frac{C}{M}}\,\,t+C_{0}\Bigr),
$$
where $C_{0}$ is an integration constant. For the case of negative
curvature ($k =-1$) and $m> 1/4$, or positive curvature ($k = +
1$) and $m <1/4$, we can obtain the following solution for
equation (\ref{15}):
$$
a(t)=
\frac{1}{\sqrt{C}}\cosh\Bigl(\sqrt{\frac{C}{M}}\,\,t+C_{0}\Bigr).
$$
where $C_{0}$ is an arbitrary constant. It is interesting that in
these cases, as it can be observed from the special feature of
solutions and equation (\ref{15}), this model behaves similarly to
the FRW model in which the only source of gravity is the effective
cosmological constant $\Lambda= 3 C/M $.

{\bf(II)} To involve  in our consideration the widely discussed
modifications of Y-M theory (see, for example,
\cite{C36}-\cite{C39}, \cite{C42}), we assume that function
$\Psi(I) =B \, I^{m} $, and  $\Phi(I)$ arbitrarily depends on $I$.
Then the main equations of our model can be written down as
follows:
\begin{equation}
\label{16}(1-4m)\Bigl[2\frac{\ddot a}{a}+\Bigl( \frac{\dot
a}{a}\Bigr)^{2} (1-4m) + \frac{k}{a^{2}}\Bigr] = \frac{8\pi G}{3
B}\,\frac{[3\Phi (I)-4\,I\,\Phi^{\prime}(I)]}{I^{m}},
\end{equation}
\begin{equation}
\label{17}\Bigl(\frac{\dot a}{a}\Bigr)^{2} (1-4m) +
\frac{k}{a^{2}}= \frac{8\pi G}{3B}\,\frac{\Phi(I)}{I^{m}}.
\end{equation}

Combining these equations, it is possible to obtain, instead of
(\ref{16}), the following equation for the second derivative of
the scale factor:
\begin{equation}
\label{18}\frac{\ddot a}{a} = \frac{8\pi G}{3 B}\,I^{-m}\, \frac
{[(1+2m)\Phi(I)-2\,I\, \Phi^{\prime}(I)]}{1-4m}.
\end{equation}
It is easy to  verify that (\ref{18}) is a differential
consequence of equation (\ref{17}) which remains the only
independent equation of our model. Nevertheless, from equation
(\ref{18}) it is possible to obtain a necessary condition for the
accelerated regime (see Table 2). \vspace{3mm}
\begin{center}
\begin{tabular}
[c]{|l||l|l|l|}\hline
\vspace{-3mm} \, & \, & \, & \,\\
IIa & $B > 0$ & $m > 1/4$ & $\Theta (I)>0$\\\hline
\vspace{-3mm} \, & \, & \, & \,\\
IIb & $B > 0$ & $m < 1/4$ & $\Theta (I)<0$\\\hline
\vspace{-3mm} \, & \, & \, & \,\\
IIc & $B < 0$ & $m > 1/4$ & $\Theta (I)<0$\\\hline
\vspace{-3mm} \, & \, & \, & \,\\
IId & $B < 0$ & $m < 1/4$ & $\Theta (I)>0$\\\hline
\end{tabular}
\end{center}
\begin{quote}
\textbf{Table 2.} Conditions for $m$ and function $\Theta (I) =
2\,I\,\Phi^{\prime}(I)-(1+2m)\Phi(I)$, corresponding to
accelerated expansion according to equation (\ref{18}).
\end{quote}
\vspace{3mm}

\noindent

Let us then consider the case of non-Abelian Lagrangian of the
Born-Infeld type (BI):
$$
L_{NBI}=\frac{\beta^{2}}{4 \pi}\Bigl(\sqrt{1+\frac{F_{ik}^{a}
F^{aik}} {\beta^{2}}-\frac{(\tilde{F}_{ik}^{a}
F^{aik})^{2}}{16\beta^{4}}}\,-1\Bigr),
$$
where $\beta$ is the critical intensity of BI-field,
$\tilde{F}_{ik}^{a}$ is a dual Y-M tensor. From formulas (\ref{4})
and metrics (\ref{2}), it follows that the second invariant of Y-M
field for our solution $\tilde{\mathbf{F}} _{ik} \mathbf{F} ^{ik}
=0$. So, we can write down $\Phi(I) $ as
\begin{equation}
\label{19}\Phi(I)=\frac{1}{16
\pi\alpha}\Bigl(\sqrt{\mathstrut1+2\alpha I}\,-1\Bigr),
\end{equation}
where $\alpha= 1/2\beta^{2}$. From the latter, it is easy to find
that
\begin{equation}
\label{20}2\,I\, \Phi^{\prime}(I)-(1+2m)\Phi(I)=\frac{1}{16\pi\alpha}%
\,\frac{[(2m+1)(\sqrt{\mathstrut1+2\alpha I}-1)-4m\alpha I]}{\sqrt
{\mathstrut1+2\alpha I}}.
\end{equation}
In view of inequalities $2 \alpha I>0$ and $\sqrt{1+2\alpha I}>
1$, one can find that expression (\ref{20}) will be positive, only
if $m\in(0,1/4) $. Thus, as it follows from the explicit form of
invariant (\ref{7}), the positivity of expression (\ref{20}) will
occur only after crossing the critical value of scale factor:
$$
a(t) > a_{cr}=\Bigl[\frac{48\,\alpha\,
m^{2}}{e^{2}(1-4m^{2})}\Bigr]^{1/4}.
$$
One can see that this case corresponds to (IId) in Table 2, if $B
<0$.

When the value of (\ref{20}) is negative, $m$ remains arbitrary.
However, the acceleration conditions  are different for $m^{2}
<1/4$ or $m^{2}> 1/4$. The first case corresponds to (IIb) with
$m\in(-1/4,1/4) $ and $B> 0$. The cosmic acceleration is possible
while $ 0<a(t)< a_{cr}$.  In the case (IIc) (i.e. for $m> 1/4,\,B
<0$), the accelerated expansion is always possible.

So non-trivial dependence of the model behavior on free parameter
$m$ requires a thorough research. Therefore, we are going to give
more details and consequences of the model considered here in our
further investigation.

\section{\large Conclusion}

\quad In summary, the modified non-minimal Y-M theory in FRW
non-flat cosmology are studied in this paper. First of all, we
have derived the set of main equations which determines the model
dynamics: (\ref{6}), (\ref{8}), (\ref{9}). Throughout the last
section of the paper, non-minimal coupling to gravity is described
by the factor $\Phi(I)$ in the Einstein-Hilbert sector of action
(\ref{1}).  The non-trivial solution of the modified Y-M equation
(\ref{8}) proposed by one of the authors (V.K.S) earlier allows us
to build several modifications of accelerated cosmic expansion in
the frame of non-minimal coupling. Besides general study, we have
considered in detail the power-law dependence of $\Phi(I)$ on its
argument. We have derived the basic equations for the cosmic scale
factor in our model, and have provided several examples of their
solutions. This work implies that the cosmological applications of
modified Y-M theory with non-minimal coupling to gravity may have
more fruitful phenomena, which is worth studying further.

\end{document}